\documentclass[journal=jacsat,manuscript=article,layout=twocolumn,articletitle=true]{achemso}  
\usepackage[T1]{fontenc}       
\usepackage{graphicx}
\usepackage{ulem}
\usepackage[linktocpage=true,colorlinks=true,pdfborder={0 0 0},linkcolor=blue,
citecolor=blue,filecolor=yellow,urlcolor=blue,bookmarks,pdfauthor={},]{hyperref}
\usepackage{gensymb}
\usepackage{amsmath}
\usepackage{multirow}
\setkeys{acs}{usetitle=true}

\newcommand{\HP}{Department of Physics and Astronomy and HiPSEC, University of Nevada Las Vegas, Las Vegas, Nevada 89154, USA}
\newcommand{\PHYSICS}{Department of Physics and Astronomy, University of Nevada Las Vegas, Las Vegas, Nevada 89154, USA}
\newcommand{\GEOLOGY}{Department of Geoscience, University of Nevada Las Vegas, Las Vegas, Nevada 89154, USA}
\newcommand{\HPCAT}{HPCAT, X-ray Science Division, Argonne National Laboratory, Argonne, Illinois 60439, USA}

\author{Zachary~M~Grande}
 \affiliation{\HP}
\author{Chenliang~Huang}
 \affiliation{\PHYSICS}
\author{Dean~Smith}
 \affiliation{\HP}
\author{Jesse~S~Smith}
 \affiliation{\HPCAT}
\author{John~H~Boisvert}
 \affiliation{\PHYSICS}
\author{Oliver~Tschauner}
 \affiliation{\GEOLOGY}
\author{Jason~H~Steffen}
 \affiliation{\PHYSICS}
\author{Ashkan~Salamat}
 \email{salamat@physics.unlv.edu}
  \affiliation{\HP}

\date{\today}

\title{Bond strengthening in dense H$_{2}$O and implications to planetary composition}

\begin{document}


\maketitle

\begin{abstract}
   \textbf{H$_{2}$O is an important constituent in planetary bodies, controlling habitability and, in geologically-active bodies, plate tectonics.
   At pressures within the interior of many planets, the H-bonds in H$_{2}$O collapse into stronger, ionic bonds.
   Here we present agreement between X-ray diffraction and Raman spectroscopy for the transition from ice-VII to ice-X occurring at a pressure of approximately 30.9\,GPa by means of combining grain normalizing heat treatment \textit{via} direct laser heating with static compression. This is evidenced by the emergence of the characteristic Raman mode of cuprite-like ice-X and an abrupt 2.5-fold increase in bulk modulus, implying a significant increase in bond strength.
   This is preceded by a transition from cubic ice-VII to a structure of tetragonal symmetry, ice-VII$_{\text{t}}$ at 5.1\,GPa.
   Our results significantly shift the mass/radius relationship of water-rich planets and define a high-pressure limit for release of chemically-bound water within the Earth, making the deep mantle a potential long-term reservoir of ancient water.
   }
\end{abstract}

The pressure-temperature phase diagram of H$_{2}$O exhibits remarkable polymorphism, with as many as 18 phases currently reported.~\cite{dunaeva2010phase,millot2019nanosecond}
At low pressures, this complexity arises from steric rearrangements of hydrogen-bonded molecules, while the H--O--H bond angle and length remain almost constant.
H-bonds are established through correlated disorder of the protons between adjacent oxygen atoms such that, at each moment, two protons and one oxygen form an H$_{2}$O molecule.~\cite{li1993evidence}
Ice structures generally exhibit network-like topologies similar to those of silica and silicates.~\cite{hazen1996high}
The behavior of condensed H$_{2}$O phases (ices) is dominated by this H-bond network.
Under the conditions found in the interior of Earth and many other planets, the H-bonds in ice are gradually replaced by ionic bonds in ice-X.~\cite{holzapfel1984effect,aoki1996infrared,goncharov1996compression}

At room temperature, ice-VII becomes the stable solid phase of H$_{2}$O at pressures above 2.7\,GPa.~\cite{bezacier2014equations} 
The subsequent transition into ice-X has been observed in spectroscopic measurements and inferred from structural data, but there has been no consensus between studies.
Estimates of the transition pressure range from 40\,GPa to above 120\,GPa.~\cite{aoki1996infrared,goncharov1996compression,wolanin1997equation,goncharov1999raman,loubeyre1999modulated,somayazulu2008situ,sugimura2008compression,guthrie2013neutron,zha2016new,meier2018observation,guthrie2019structure}
However, the existence of a molecular-to-ionic transition above 40\,GPa has been observed at temperatures beyond the melting curve.~\cite{schwager2008h2o,goncharov2009dissociative}

Bond states in soft molecular compounds are strongly affected by non-hydrostatic stress at high pressure conditions.
H$_{2}$O is especially susceptible to this since the use of a pressure transmitting medium is inhibited due to the formation of hydrates and clathrates.
The resulting distortions caused by non-hydrostatic compression are further exacerbated by the heterogeneous nucleation of ice-VII within ice-VI, which yields large crystalline domains and causes significant anisotropic effects at grain boundaries.~\cite{Stern_1997_grain}

To minimize these effects, we heat ice samples at high pressure using a CO$_{2}$ laser and allow them to cool to ambient temperature.
The cooling rate is slower than rapidly quenching, which can potentially trap internal stresses, and is faster than annealing, which typically results in enlarged domains (see Methods and Supplementary Information), and an analogy can be made with metallurgical normalization.~\cite{thelning2013steel}

There are several benefits of this heat treatment:
anisotropic strain within both the sample and the Au pressure marker are relieved, minimizing deviatoric stress for more accurate volume-pressure measurements; the recrystallization of the ice produces a powdered sample with nanoscopic domains in random orientation (Supplementary~Video); and provides a direct, localized method of heating.
The reduced domain size and their random orientations yield well-resolved Debye-Sherrer rings for an extensive $q$-range (Figure~\ref{fig:BFplot}a), making our data suitable for Rietveld powder X-ray diffraction analysis (Figure~\ref{fig:BFplot}b).
The powdered nature of the sample also reduces its susceptibility to further strain as compression continues, despite the uniaxial nature of the DAC, allowing full structural refinement.
Data that are not heat treated display significantly fewer diffraction features and typically exhibit multi-grain spots or highly textured rings with significant peak broadening from deviatoric strain, shown in Figure~\ref{fig:BFplot}a where the FWHM of the (1\,1\,0) peak improves from 0.24 to 0.088\degree{}\,2$\theta{}$ in the heat treated pattern.

\section{Results}

\begin{figure*}
\includegraphics[]{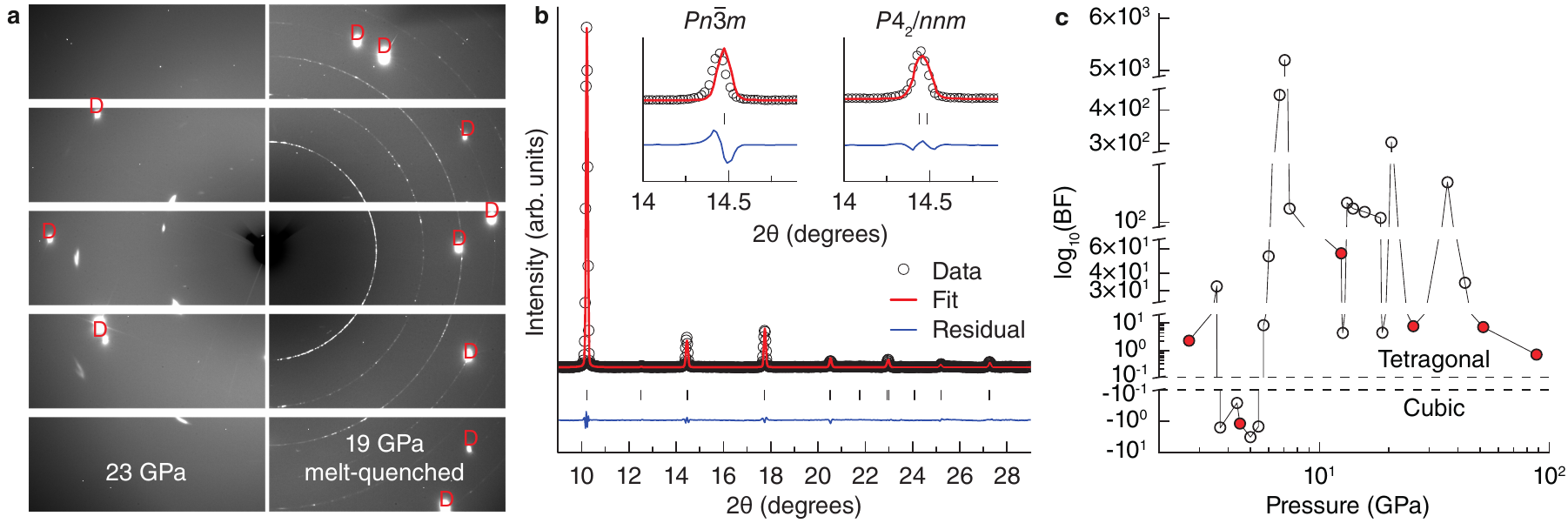}
\caption{\textbf{Tetragonal distortion of ice-VII.}
\textbf{a} Comparison of raw XRD images of ice: (left) highly strained and textured diffraction pattern at 23.2\,$\pm{}$\,0.6\,GPa without heat treating and (right) full Debye-Sherrer rings from heat treating an annealed powder of ice at 19.1\,$\pm{}$\,0.4\,GPa.
Each red letter D indicates a reflection from the single-crystal diamond anvil.
\textbf{b} Rietveld refinement of ice-VII$_{\text{t}}$ $(P4_{2}/nnm)$ at 6.5\,$\pm{}$\,0.5\,GPa ($a$\,=\,3.2279\,$\pm{}$\,0.0002\,\AA{} and $c$\,=\,3.2372\,$\pm{}$\,0.0003\,\AA{}, $V$\,=\,33.719\,$\pm{}$\,0.002\,\AA{}$^{3}$, $wR_{\text{P}}$\,=\,1.81\% and R$_{\text{P}}$\,=\,1.41\%).
Inset: (left) Rietveld refinement of the (2\,0\,0) Bragg peak using a cubic cell  ($Pn\overline{3}m$) ($a$\,=\,3.2275\,$\pm{}$\,0.0002\,\AA{}, $V$\,=\,33.621\,$\pm{}$\,0.005\,\AA{}$^{3}$, $wR_{\text{P}}$\,=\,2.36\% and R$_{\text{P}}$\,=\,2.43\% ) and (right) improved fit using a tetragonal cell $(P4_{2}/nnm)$ (2\,0\,0) and (0\,0\,2) Bragg peaks.
\textbf{c} Log Bayes Factors (BFs) for a tetragonal model \textit{vs} cubic model for our data on a logarithmic scale.
Positive values indicate that the data favor the tetragonal model where negative values indicate preference for the cubic model.
Red points indicate pressures that the sample was laser heated.  We used a random sampling of points across the pressure range of our experiment to conduct this analysis.
}
\label{fig:BFplot}
\end{figure*}

A single-phase sample of powdered ice-VII is achieved by heating at the first appearance of its coexistence with ice-VI, as confirmed with XRD measurements (See Methods and Supplementary Information).
Beginning at 2.7\,$\pm{}$\,0.4\,GPa, we unambiguously index and refine the phase as cubic ice-VII.
Above 5.1\,$\pm{}$\,0.5\,GPa, we observe deviations in the peak positions and profiles both before and  -- more clearly -- after heat treating.
Specifically, we observe splitting between the (2\,0\,0)/(0\,0\,2) which are not accommodated by the cubic ice-VII ($Pn\overline{3}m$) structure.
Figure~\ref{fig:BFplot}b shows the Bragg feature at $\sim{}$\,14.5\degree{}, where these deviations are most pronounced.
Here, we find significant improvements in the Rietveld refinement when modelling with a tetragonal sub-group of cubic ice-VII ($P4_{2}/nnm$), and name this tetragonal phase ice-VII$_{\text{t}}$.

Structural anomalies have been reported in the 10--14\,GPa regime,~\cite{wolanin1997equation,somayazulu2008situ,guthrie2013neutron,okada2014electrical} and are attributed to a proton-disordered ice-VII$^{\prime{}}$, but these claims have lacked spectroscopic evidence.
One such anomaly has been suggested to result from a tetragonal distortion.~\cite{somayazulu2008situ}
We show that this transition from ice-VII to ice-VII$_{\text{t}}$ is accompanied by a 2.18\,$\pm{}$\,0.01\% volume collapse of the unit cell at this pressure (Supplementary Information).
We only confirm a transition in the oxygen sublattice to a tetragonal symmetry and cannot comment on its relation to proton-disordered ice-VII$^{\prime{}}$, as the few observations of the (1\,1\,1) diffraction peak were dominated by the background in our XRD patterns.

We further examine the symmetry of the unit cell above $\sim{}$5\,GPa, by combining structural refinements with a Bayesian model comparison algorithm (see Methods).~\cite{Boisvert18}
Figure~\ref{fig:BFplot}c shows the log Bayes factors comparing cubic and tetragonal models for a selection of points over the pressure range of our data.
A cubic model is favoured for pressures below 5\,GPa, indicated by Bayes factors below unity. 
Meanwhile, the tetragonal model is clearly preferred between 5 and 30\,GPa, shown by values much larger than unity.
Above 30\,GPa, the cubic model becomes increasingly viable -- consistent with our Reitveld refinements, as well as previous diffraction-based studies.~\cite{loubeyre1999modulated, sugimura2008compression}.
We find no static displacement of the O-sublattice in tetragonal ice-VII$_{\text{t}}$, suggesting that the gradual softening of the O--H vibrational mode over its stability range~\cite{holzapfel1984effect,goncharov1996compression,aoki1996infrared} is solely related to the weakening of the H-bond.

\begin{figure*}
\includegraphics[]{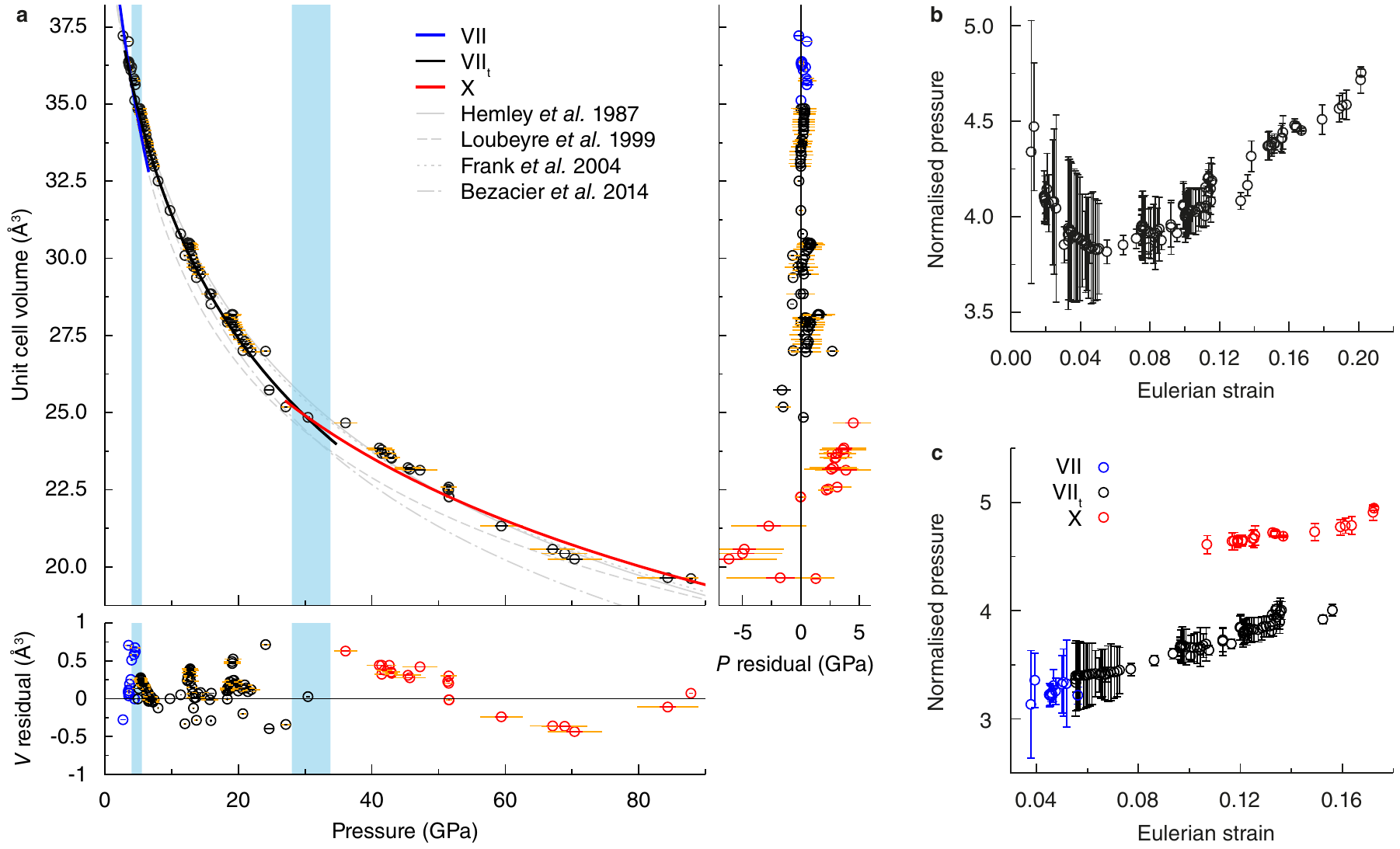}
\caption{
\textbf{Equation of state fitting.}
\textbf{a} Pressure-volume plot of our data and Vinet EOS fit from MCMC for the three phases.
The calculated uncertainty in transition pressures are indicated by the blue regions at 5.1\,$\pm{}$\,0.5\,GPa and 30.9\,$\pm{}$\,2.9\,GPa, respectively, and the grey lines are results from previous experiments.
Curves are colour coded by phase
(blue: cubic ice-VII ($K_{0}$\,=\,18.47\,$\pm{}$\,4.00\,GPa, $K_{0}^{\prime{}}$\,=\,2.51\,$\pm{}$\,1.51, $V_{0}$\,=\,42.50\,$\pm{}$\,0.88\,\AA{}$^{3}$),
black: non-cubic ice-VII$_{\text{t}}$ ($K_{0}$\,=\,20.76\,$\pm{}$\,2.46\,GPa, $K_{0}^{\prime{}}$\,=\,4.49\,$\pm{}$\,0.35, $V_{0}$\,=\,41.11\,$\pm{}$\,0.53\,\AA{}$^{3}$),
and red: ice-X ($K_{0}$\,=\,50.52\,$\pm{}$\,4.16\,GPa, $K_{0}^{\prime{}}$\,=\,4.50\,$\pm{}$\,0.15, $V_{0}$\,=\,33.82\,$\pm{}$\,0.43\,\AA{}$^{3}$)).
In comparison, fitting all of the data to a single phase yields $K_{0}$\,=\,12.57\,$\pm{}$\,0.50\,GPa, $K_{0}^{\prime{}}$\,=\,6.06\,$\pm{}$\,0.07, and $V_{0}$\,=\,43.05\,$\pm{}$\,0.20\,\AA{}$^{3}$.
Orange error bars indicate our systematic uncertainty from deviatoric stresses in non heat treated data.
\textbf{b} Linearized Vinet EOS when applying a single phase.
\textbf{c} Three-phase linearized Vinet EOS.
}
\label{fig:eos}
\end{figure*}

We perform a similar Bayesian analysis comparing a single-phase equation of state (EOS) to a three-phase EOS and find that a three-phase model is required to reproduce our data.
In doing so, we fit a three-phase $P$-$V$ Vinet EOS to the data using Markov Chain Monte Carlo (MCMC).
This model includes two transition pressures as fitting parameters as well as two parameters, $\beta{}$ and $\gamma{}$, to model pressure-dependent systematic uncertainties (Supplementary Information). 
The uncertainties for data that were not heat treated are adjusted by the function, $\sigma{}=\sigma_{0}(\beta{}+\gamma{}P)$ (orange error bars in Figure~\ref{fig:eos}a), while heat treated data (where distortions in our Au pressure marker are relieved) use their nominal error, $\sigma_{0}$. 
The results of our three-phase fit of the $P$-$V$ EOS and the transition pressures between the phases is shown in Figure~\ref{fig:eos}a.
The Bayes factor comparing the three-phase fit to the single-phase fit is 3.21$\times\,10^{86}$ -- strongly favoring the three-phase model (See Supplementary Information for likelihoods and priors).

This three-phase model is best highlighted with the linearized form of the Vinet EOS relating the normalized pressure to Eulerian strain,~\cite{angel2000equations} which is sensitive to the starting phase volume, $V_{0}$.  
Using $V_{0}$ from a single-phase fit to ice-VII fails to describe the compressibility across the entire pressure range as seen by the abnormal curvature in Figure~\ref{fig:eos}b.
Conversely, linear trends appear when modelling with three distinct phases (Figure~\ref{fig:eos}c). 

\begin{figure*}
    \centering
    \includegraphics[]{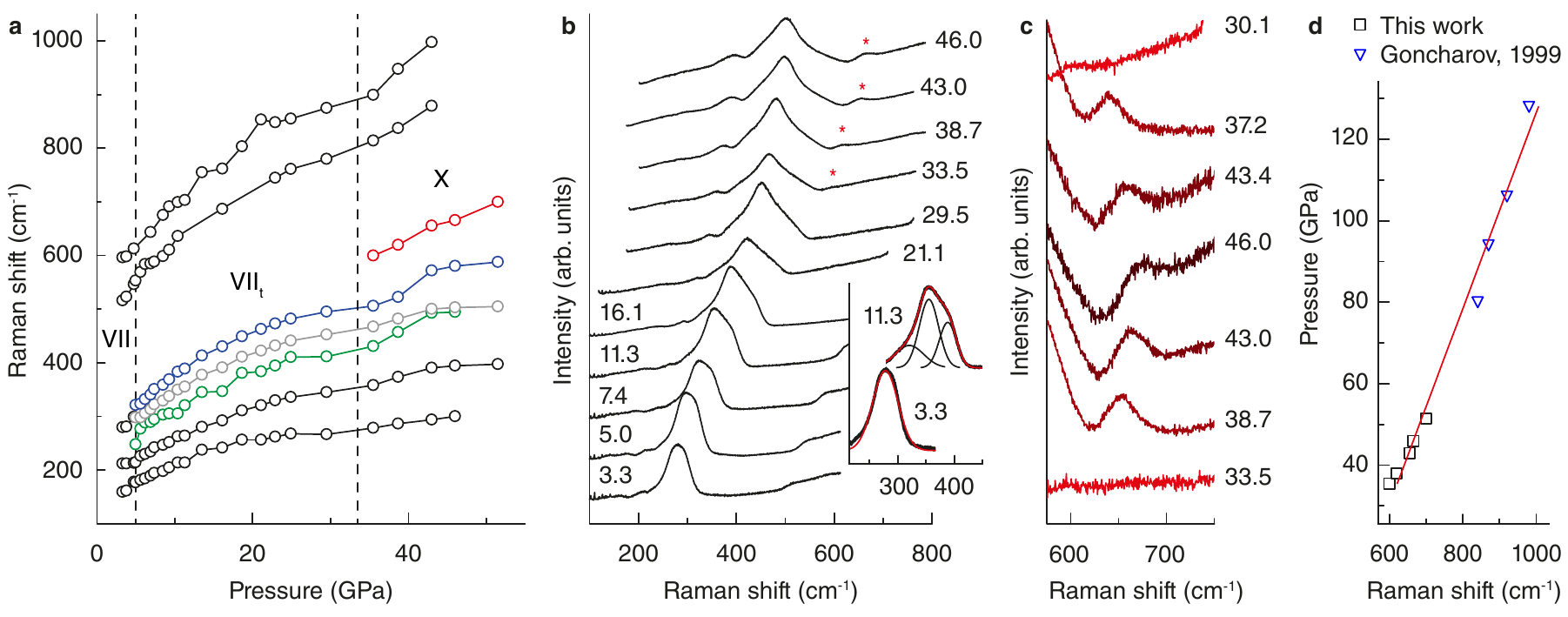}
    \caption{\textbf{Raman spectrum of heat treated H$_{2}$O ices under compression.}
    \textbf{a} Frequency shift of measured Raman modes of H$_{2}$O ice with pressure.
    Splitting of the dominant lattice mode near 280\,cm$^{-1}$ due to tetragonal distortion above 5\,GPa is highlighted in blue and green.
    Red diamonds show the emergence of the ice-X $T_{\text{2g}}$ mode above 33\,GPa. Dashed lines represent transition pressures based on analysis of XRD data.
    \textbf{b} Progression of Raman features on increasing pressure.
    The dominant mode near 280\,cm$^{-1}$ exhibits asymmetry above 5\,GPa, and tends towards a single mode above 21\,GPa.
    Red asterisks (\textcolor{red}{*}) denote the emergence of ice-X $T_{\text{2g}}$ Raman mode.
    (Inset) Dominant Raman feature in ice-VII at 3.3\,GPa is fit to a single peak, whereas the same feature in ice-VII$_{\text{t}}$ at 11.3\,GPa is a triplet.
    \textbf{c} (bottom-to-top) Development of ice-X $T_{\text{2g}}$ mode on compression above 33\,GPa, and its reversible disappearance on decompression.
    \textbf{d} Frequency shift of ice-X $T_{\text{2g}}$ Raman mode with pressure, showing correlation with those reported by \citet{goncharov1999raman} at higher pressures.
    }
    \label{fig:raman}
\end{figure*}

Additionally, we confirm both the tetragonal structure near 5\,GPa and onset of ice-X near 30\,GPa through Raman spectroscopy.
We first observe the lattice modes of ice-VII near 3.3\,GPa after heat treatment of the solidified sample (Figure~\ref{fig:raman}a).
This spectrum is dominated by a feature at 280\,cm$^{-1}$ which is fit to a single mode and a weaker mode at 211\,cm$^{-1}$ (Figure~\ref{fig:raman}a and b inset). 
Due to the close similarities between the Raman spectra of proton-disordered ice-VII and its proton-ordered analogue, ice-VIII, these modes have previously been assigned to the analogous translational-vibration modes of ice-VIII, $B_{\text{1g}}$ and $A_{\text{1g}}$, respectively.~\cite{pruzan1990raman,zha2016new}.
We also observe a very weak mode which is not reported previously near 160\,cm$^{-1}$ (Supplementary Information).
Features were also observed in the 500\,cm$^{-1}$ to 800\,cm$^{-1}$ range corresponding to the known $E_{\text{g}}$ and $B_{\text{2g}}$ rotational modes.\cite{zha2016new}

Beginning at approximately 5.0\,GPa, the dominant feature near 280\,cm$^{-1}$ displays an increasingly asymmetric profile requiring multiple modes to reproduce the peak profile (Figure~\ref{fig:raman}a and b inset).
This asymmetry arises due to the appearance of new lattice modes (Figure~\ref{fig:raman}b inset) which is consistent with a lowering of symmetry from the cubic $Pn\overline{3}m$ space group  to the tetragonal $P4_{2}/nnm$ space group.
This asymmetric profile continues with further compression and at approximately 21.1\,GPa, the peak profile becomes increasingly symmetric (Figure~\ref{fig:raman}b), mirroring the results obtained from the Bayesian model comparison of our XRD data shown in Figure~\ref{fig:BFplot}c.

By 38.7\,GPa, we observe a new feature at 618\,cm$^{-1}$ (Figure~\ref{fig:raman}b), the intensity of which increases with pressure, and is observed to disappear upon decompression with little hysteresis (Figure~\ref{fig:raman}c).
We interpret this new mode as the $T_{\text{2g}}$ mode, signifying the onset of ice-X.\cite{hirsch1986effect,goncharov1999raman}
Furthermore, there is a noticeable stiffening of the the frequency of the lattice modes, shown in Figure~\ref{fig:raman}a, following the emergence of this new peak which agrees with our observation of bond strengthening at the onset of ice-X in this region.
The emergence of a new mode correlates with our calculated transition pressure to ice-X based on the Bayesian analysis of our equation of state data (Figure~\ref{fig:eos}a), as well as the observations of \citet{hirsch1986effect}, in which the transition from ice-VIII to ice-X is accompanied by the emergence of the $T_{\text{2g}}$ mode -- the only Raman active mode of cuprite-like ice-X. 
Due to the aforementioned similarities between the Raman spectra of ice-VII and ice-VIII,~\cite{pruzan1990raman,zha2016new} this agreement supports our claim of a transition to ionic-bonded ice-X in this region.
The frequency of this mode is tracked with pressure up to 51.5\,GPa, and is in good agreement with measurements made by \citet{goncharov1999raman} above 80\,GPa as shown in Figure~\ref{fig:raman}d.

\section{Discussion}

\begin{figure}[]
\includegraphics[width = \columnwidth{}]{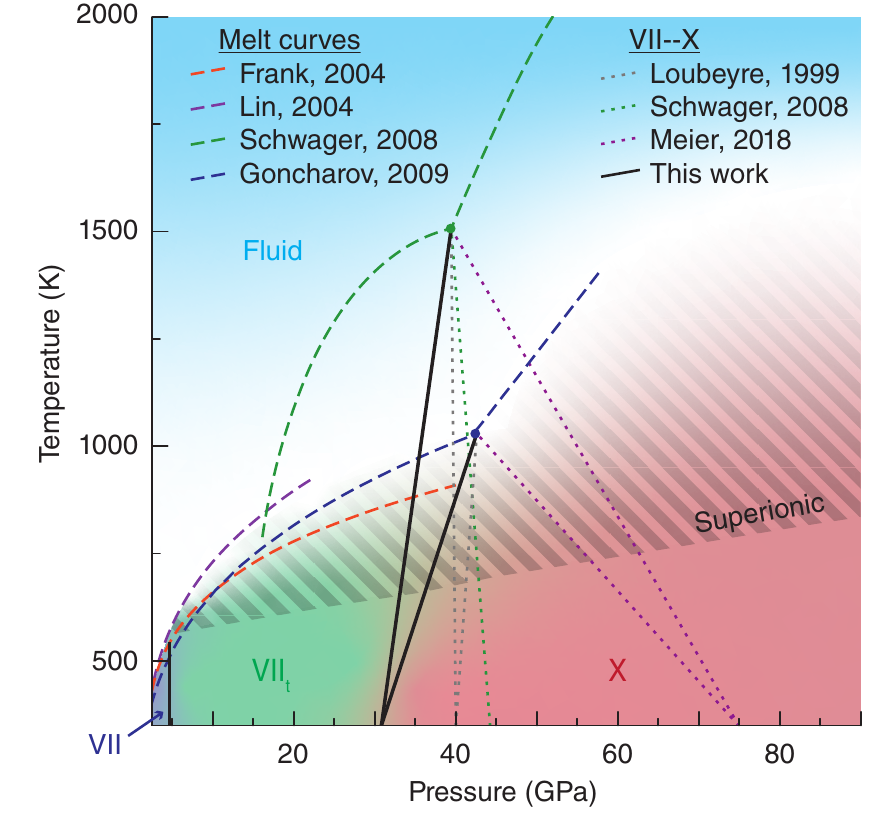}
\caption{\textbf{High-pressure high-temperature phase diagram of H$_{2}$O.}
Dark blue, green and red shaded regions denote ice-VII, VII$_{\text{t}}$ and X, respectively, and projected phase boundaries separating high-pressure ice phases from our work are shown as solid black lines.
Ice-X phase boundaries connect our measured transition at 30.91\,$\pm{}$\,2.90\,GPa and 300\,K to the inflection point in the melt curve observed by \citet{schwager2008h2o} and \citet{goncharov2009dissociative}, which have been associated with the transition from molecular to ionic fluid.
The same procedure has been used to project phase boundaries from \citet{loubeyre1999modulated} and \citet{meier2018observation}
In doing so, we deduce a steep, positive Clapeyron slope defining the transition from hydrogen bonding to ionic bonding in dense H$_{2}$O, consistent with a pressure-driven change in bonding nature.
Dashed lines show measured melting curves.~\cite{Frank2004,Lin_2004_melt,schwager2008h2o,goncharov2009dissociative}
Superionic boundary from~\citet{sugimura2012experimental} is highlighted.
}
\label{fig:phasediagram}
\end{figure}

The results of our multi-phase fit show that room temperature H$_{2}$O takes the form of cubic ice-VII from 2.7\,$\pm{}$\,0.4 to 5.1\,$\pm{}$\,0.5\,GPa, followed by tetragonal ice-VII$_{\text{t}}$ to 30.9\,$\pm{}$\,2.9\,GPa, then cubic ice-X thereafter (Figure~\ref{fig:phasediagram}).
These transitions are similarly observed in our Raman study and marks the first agreement between XRD and spectroscopy for the transition from ice-VII to ionic-bonded ice-X.
The low transition pressure into non-cubic ice-VII$_{\text{t}}$ based on an observed lowering of symmetry, implies that cubic ice-VII is stable for only a small window of phase space -- contrary to existing assumptions~\cite{holzapfel1972symmetry}, making ice-VII$_{\text{t}}$ the most abundant phase of ice in the crust and upper mantle of water-rich planets.
At 30.9\,$\pm{}$\,2.9\,GPa, a 2.5-fold increase in bulk modulus marks a significant strengthening of the O--H--O bond.~\cite{born1954dynamical} We interpret this bond strengthening as the transition from H-bonded ice-VII$_\text{t}$ to ionic-bonded ice-X as is supported by the emergence of the $T_{\text{2g}}$ mode.


There is an abrupt steepening of the melt curve of H$_{2}$O near 44\,GPa,~\cite{schwager2008h2o,goncharov2009dissociative} which must result from a change in physical properties either within the fluid phase above, or the solid phases below (Figure~\ref{fig:phasediagram}).
Our observed abrupt transition to ice-X near 30\,GPa is consistent with this increase in the slope of the melting temperature around 44\,GPa, and the occurrence of superionic ice as an intermediate phase between ice-X and melt at around 50\,GPa.~\cite{sugimura2012experimental,millot2018experimental}
Assignment of the inflection in the melting line to the onset of ice-X above 30\,GPa results in a steep, positive Clapeyron slope at the phase boundary as shown in Figure~\ref{fig:phasediagram}.
Our results suggest that the transition from H-bonded to ionic H$_{2}$O (solid black lines in Figure~\ref{fig:phasediagram}) is not strongly temperature-dependent, and that the change in bonding in high-pressure H$_{2}$O ices is primarily pressure-driven, consistent with the observation of ionic fluids at similar pressures and high temperatures.~\cite{goncharov2009dissociative}

These results have fundamental consequences for the global water cycle. 
Most chemically-bound water on Earth is released from subducted slab at shallow depths in the upper mantle, transition zone, and the shallow lower mantle.~\cite{schmandt2014dehydration,tschauner2018ice}
However, our data show that H$_{2}$O in the ionic-bonded pressure regime will be significantly less compressible than in the H-bonded regime. 
This has the effect of stabilising chemically-bound water in lower mantle minerals,~\cite{Walter_2015_hydrous} inhibiting fluid release at greater depths. 
Thus, residual water that is chemically bound in the deep lower mantle will remain trapped, making this region of the Earth a reservoir of ancient water that is possibly only released by mantle plumes.~\cite{hallis2015evidence}

\begin{figure*}[!hbp]
\includegraphics[width=\textwidth]{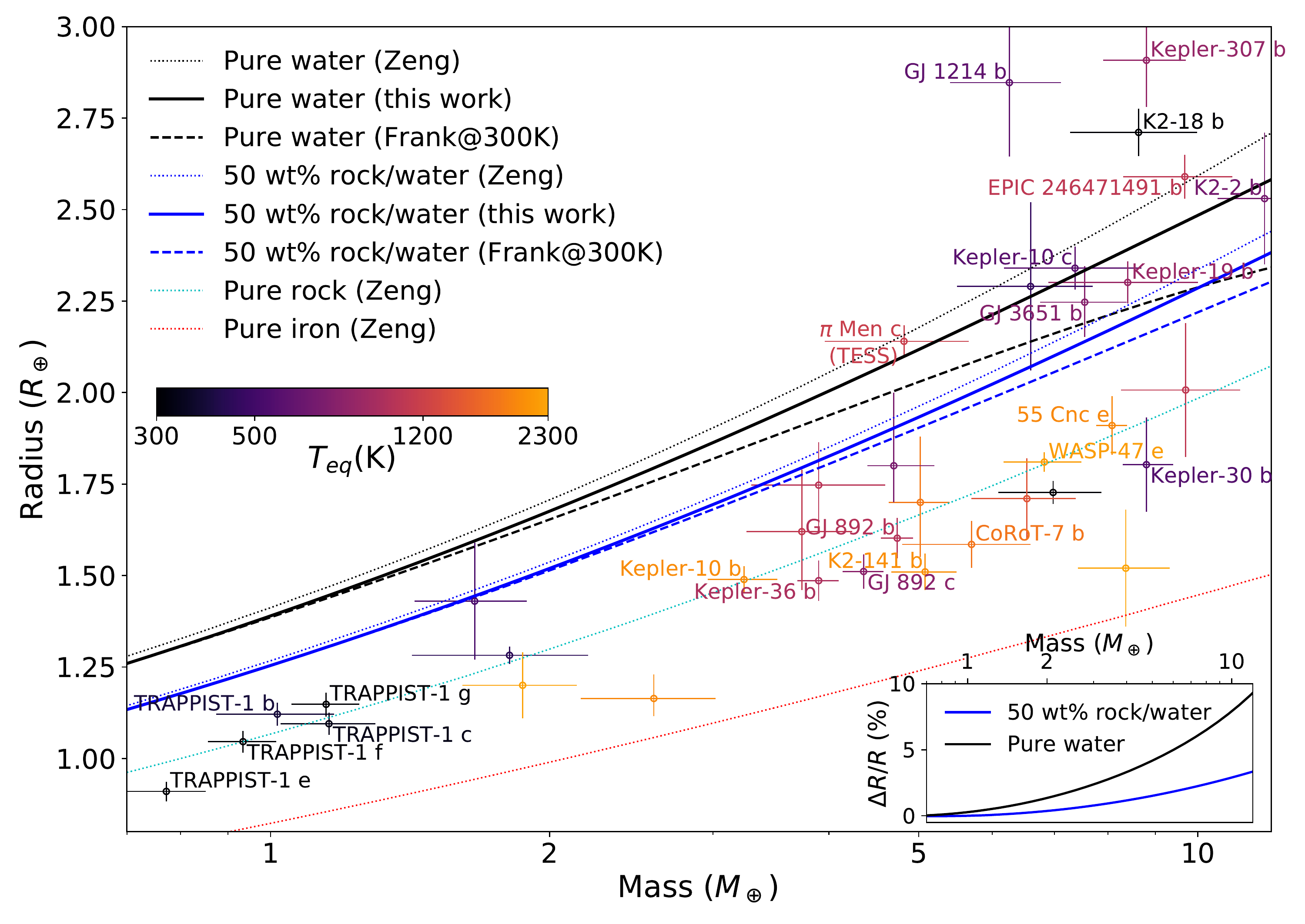}
\caption{
\textbf{Mass-radius curve of planet models compared to observational data.}
Results from the three-phase EOS in this work are shown by the solid mass-radius curves.
Dotted curves show the result from \citet{Zeng2016}, which considered the effects of planet interior temperature variations.  
The dashed curves show a more direct comparison to our results by removing the temperature dependence by substituting the high-pressure phases of ice in their model with a layer of ice-VII at 300~K using the EOS from \citet{Frank2004} (Methods).  
Planets whose radii and masses are measured to better than $\sim{}10\%$ and $\sim{}20\%$ respectively are plotted and colour-coded by their surface temperatures.
Source: NASA Exoplanet Archive, TEPCat\cite{TEPCat} and thereafter (see Supplementary information). 
The increased planet radius suggested by the new EOS is larger than the radius uncertainty of many planets.
Thus, the contribution of the atmosphere or of water content to the planet structure for planets such as $\pi$ Mensae-c, Kepler-10c, and EPIC-246471491b may be less than previously inferred.  
(Inset) The percent difference in radius between our three-phase EOS and \citet{Frank2004}.
}
\label{RM}
\end{figure*}

The observed bond strengthening of ionic H$_{2}$O also affects the assessment of the inferred composition and structure of water-rich planets.~\cite{Zeng2016}
Planetary constituents in the crust and upper mantle contribute most significantly to a planets mass/radius profile due to their lower density and higher susceptibility to changes in pressure and temperature.  
The greater incompressibility of ice-X produces larger planets for a given mass, thereby either reducing the atmospheric contribution to the volume of many exoplanets or limiting their water content. 
For example, the amount of water in planets such as Kepler-19b and GJ-3651b must be less than previous results allow.
While planets such as EPIC-246471491b and K2-18b cannot be modelled with condensed material alone and must have substantial atmospheres.
The mass-radius relationship in Figure~\ref{RM} shows that the systematic differences between previous EOS measurements and our own are comparable to uncertainties in planet sizes.
Thus, ongoing improvements in exoplanet mass and radius measurements require similar improvements in the EOS of planetary materials -- such as those presented here -- to understand their bulk composition, structure, and internal dynamics.

\section{Methods}

\textbf{Equation of state data collection.}
We perform equation of state measurements using a diamond anvil cell (DAC) of custom design, driven by a gas membrane.
Diamond culet sizes typically range from 100--300\,$\mu{}$m.
H$_{2}$O (electrophoresis and spectroscopic grade; Sigma-Aldrich) is loaded into sample chambers formed by laser micromachining~\cite{Hrubiak_laserdrill_2015} in the liquid phase, cooled slightly to avoid risk of evaporation from the diamond surface, with a $\sim{}$10\,$\mu{}$m piece of polycrystalline Au to serve as a pressure marker.
The risk of unwanted chemistry on laser heating or the formation of clathrates~\cite{marshall1964hydrates} guides our choice to not include a pressure transmitting medium in our experiments.
Instead, laser heating was employed to directly anneal residual stresses in the sample.~\cite{Boehler_review_2000,uts2013effect}

We perform powder X-ray diffraction at the HPCAT diffraction beamline (Sector 16-ID-B, Advanced Photon Source, Argonne National Laboratory, IL, USA) using a monochromatic beam with wavelength $\lambda{}$\,=\,0.406626\,\AA{}.
Two-dimensional diffraction patterns are integrated into conventional one-dimensional spectra with the \textsc{Dioptas} software package,~\cite{Prescher_DIOPTAS_2015} and Rietveld refinements performed using \textsc{GSAS}.

\textbf{Heat treating by CO$_{2}$ laser.}
To prepare powdered water ice under high pressure, we utilize its high absorbance in the mid-infrared.
10.6\,$\mu{}$m radiation from a Synrad~\textit{Evolution}125 CO$_{2}$ laser is focused to a minimum spot size $\sim{}$\,30\,$\mu{}$m through the diamond anvil and directly onto the compressed sample, using an instrument built in-place at the HPCAT diffraction beamline (Sector 16, Advanced Photon Source, Argonne National Laboratory, IL, USA),~\cite{Smith_CO2_2018} or on a system housed at UNLV.
Visible imaging confirms the formation of a melt "bubble" within the single- or few-crystal samples, surrounded by dynamically recrystallizing powdered ice (Supplementary~Video).
The focused beam is translated throughout the sample chamber -- in this way, both the powder and the Au pressure marker are thoroughly annealed.
At higher pressures (>\,50\,GPa), a melt is not always achieved.
In these cases, CO$_{2}$ laser annealing is employed to anneal a powder prepared at lower pressures -- observable in the reduction of Bragg peak widths.
XRD measurements were taken once the sample cooled to 300\,K.

Heterogeneous nucleation of ice-VII in ice-VI has a tendency of growing large domains with a preferred orientation, creating anisotropic stress and shearing at grain boundaries, and resulting in a great degree of texturing and broadening of Bragg peaks in XRD.~\cite{balzar1993voigt}
The onset of ice-VII is thus determined by heat treating immediately upon its coexistence with ice-VI -- fine control of pressure from the membrane-driven DAC allows incremental compression until the onset of phase coexistence is evident from XRD.
The heat treated sample comprises ice-VII only, and we begin measurements on the phase at the very beginning of its field of thermodynamic stability.
This gives us accurate determination of the starting volume, $V_{\text{P}}$, of ice-VII at the transition pressure.

To avoid reactions between the heated H$_{2}$O and the Re gasket which can cause catastrophic gasket failure, we line the inside of our sample chamber with Pt.
Polycrystalline Pt is loaded inside a prepared Re gasket, compressed flat, and a new sample chamber drilled to leave only a thin ring of Pt isolating the heated sample from the Re (Supplementary Information).

\textbf{Raman spectroscopy}
We perform Raman spectroscopy on a home-built system using the 514.5\,nm line of an Ar-ion laser and a $f$/9 Princeton spectrometer employing OptiGrate filters for near-Rayleigh measurements.
Loadings were performed as in equation of state experiments.

\textbf{Unit cell and EOS model comparison using MCMC.}
Starting values for our MCMCs are determined by maximizing the likelihood function.
See the Supplementary Information for the likelihood functions.
The posterior distributions of the model parameters using an ensemble sampler MCMC.~\cite{DFM13}
See the Supplementary Information for the priors used.
Each run has a number of Markov chains equal to five times the number of model parameters (plus one if the result is odd), thins the chains every hundred steps, and ignores the first 20\% of the chain as burn-in.
The chains evolve until they yield a set of at least ten thousand independent samples per model.
The chains are initialized using the parameter values from a maximum likelihood estimate, with each parameter scattered by a sufficiently small amount to allow the ensemble sampler to fill the posterior mode.
The resulting posterior distributions yield accurate, correlated errors on the model parameters.

Finally, our procedure quantitatively compares models by calculating the Bayes factor, the ratio of the probabilities of the data given two competing models.
The Bayes factor is estimated from the ratio of their fully marginalized likelihoods (FML, \textit{i.e.} Bayesian evidence) and accounts for different numbers of model parameters.
The FML is approximated using an importance sampling algorithm where the sampling distribution is informed by a set of posterior samples taken from the aforementioned MCMC.~\cite{Nelson16,Boisvert18}

\textbf{Planet mass-radius model.}
Following the treatment in \citet{Zeng2016}, we consider a fully differentiated spherical symmetric planet composed of a rocky core and an ice shell.
For the rocky core, we continue to use their Preliminary Reference Earth Model extrapolated EOS~\cite{Zeng2016} in solving the hydrostatic equation.
For the ice shell, the third-order Vinet three-phase ice EOS is adopted above the phase transition pressure from ice-VI to ice-VII at 2.1~\,GPa.~\cite{dunaeva2010phase}
ice-X EOS is extrapolated to the pressure above 88~\,GPa.
The Vinet EOS is chosen because it extrapolates better to high pressures than the third-order Birch-Murnaghan EOS.~\cite{Cohen2000}
In planet models we considered here, the pressure of the ice can reach up to $\sim{}$\,700\,GPa at the center of a 10 Earth mass pure water planet.
Shown in the dotted line in Figure~\ref{RM}, \citet{Zeng2016} used the derived EOS along the melting line for 2.2\,$\leq{}P\leq$\,37.4\,GPa,~\cite{Frank2004} and an interpolated EOS from quantum molecular dynamics simulations at a series of discrete pressure and temperature points for 37.4\,GPa\,$\leq{}P\leq{}$\,8.89\,TPa.~\cite{French2009}.  
To better show the affect of the ice measurement on planet structure and avoid the impact of uncertain planet inner temperature profile, we reproduce the model of \citet{Zeng2016}, replacing their EOS of ice above 2.2\,GPa with the EOS of ice-VII at 300\,K that is also given by \citet{Frank2004}, shown as the dashed line in Figure~\ref{RM}. 

The EOS of water in the layer with $P<{}$\,2.1\,GPa is less important to the planet radius because of its small thickness.
In the model, according to the water phase diagram at 300~K,~\cite{dunaeva2010phase} liquid H$_{2}$O EOS~\cite{Valencia2007} applied up to 0.99\,GPa and then switch to ice-VI EOS.~\cite{bezacier2014equations}

\section*{Acknowledgements}
This research was sponsored in part by the National Nuclear Security Administration under the Stewardship Science Academic Alliances program through DOE Cooperative Agreement \#DE-NA0001982.
This work was performed at HPCAT (Sector 16), Advanced Photon Source (APS), Argonne National Laboratory.
J.S.S. acknowledge the support of DOE-BES/DMSE under Award DE-FG02-99ER45775.
HPCAT operation is supported by DOE-NNSA under Award No. DE-NA0001974, with partial instrumentation funding by NSF.
O.T acknowledges support from the National Science Foundation under award NSF-EAR 1838330.
J.H.S., J.H.B, and C.H. acknowledge support from NASA grants NNX16AK32G and NNX16AK08G.
This research made use of the Cherry Creek computer cluster administered by the UNLV National Supercomputing Institute, and the NASA Exoplanet Archive, which is operated by the California Institute of Technology, under contract with the National Aeronautics and Space Administration under the Exoplanet Exploration Program.\\

\section*{Author contributions statement}
Z.M.G., D.S., J.S.S. and A.S. conducted all the experiments, Z.M.G., J.H.B. and A.S. analysed the data.
C.H. and J.H.S. modeled planet structures, Z.M.G. wrote the manuscript.
O.T., J.H.S. and A.S. conceived the project.
All authors contributed to the discussion of the results and revision of the manuscript.


\providecommand{\latin}[1]{#1}
\makeatletter
\providecommand{\doi}
  {\begingroup\let\do\@makeother\dospecials
  \catcode`\{=1 \catcode`\}=2 \doi@aux}
\providecommand{\doi@aux}[1]{\endgroup\texttt{#1}}
\makeatother
\providecommand*\mcitethebibliography{\thebibliography}
\csname @ifundefined\endcsname{endmcitethebibliography}
  {\let\endmcitethebibliography\endthebibliography}{}

\end{document}